\documentclass[runningheads,a4paper]{llncs}

\usepackage{amssymb}
\usepackage{graphicx,epstopdf}
\usepackage{xcolor}

\usepackage{url}
\urldef{\mailsa}\path|laurent.reynaud@orange.com|
\urldef{\mailsb}\path|Isabelle.Guerin-Lassous@ens-lyon.fr|    
\newcommand{\keywords}[1]{\par\addvspace\baselineskip
\noindent\keywordname\enspace\ignorespaces#1}

\begin{document}

\mainmatter  % start of an individual contribution

\title{Physics-Based Swarm Intelligence for Disaster Relief Communications}

\titlerunning{Physics-Based Swarm Intelligence for Disaster Relief Communications}

\author{Laurent Reynaud$^{1,2}$ \and Isabelle Gu\'{e}rin-Lassous$^2$}
\authorrunning{Physics-Based Swarm Intelligence for Disaster Relief Communications}

\institute{$^1$Orange Labs, Lannion, France \\
$^2$Universit\'{e} de Lyon / LIP (ENS Lyon, CNRS, UCBL, INRIA), Lyon, France\\
\mailsa\\
\mailsb\\
}

\toctitle{Physics-Based Swarm Intelligence for Disaster Relief Communications}
\tocauthor{Reynaud and Gu\'{e}rin-Lassous}
\maketitle

\begin{abstract}
This study explores how a swarm of aerial mobile vehicles can provide network connectivity and meet the stringent requirements of public protection and disaster relief operations. In this context, we design a physics-based controlled mobility strategy, which we name the extended Virtual Force Protocol (VFPe), allowing self-propelled nodes, and in particular here unmanned aerial vehicles, to fly autonomously and cooperatively. In this way, ground devices scattered on the operation site may establish communications through the wireless multi-hop communication routes formed by the network of aerial nodes.
We further investigate through simulations the behavior of the VFPe protocol, notably focusing on the way node location information is disseminated into the network as well as on the impact of the number of exploration nodes on the overall network performance.

\end{abstract}

% =============
% Section: introduction
% =============
\section{Introduction}

In the wake of disaster, communication systems play a key role for the support of appropriate response operations. During such times, when existing terrestrial  networks may be damaged or even completely impaired, Public Protection and Disaster Relief (PPDR) teams need to lean on a reliable emergency communications infrastructure to be able to restore essential services and more generally to quickly provide an adequate assistance to the affected population~\cite{cept2}. Regarding these requirements, multiple temporary network architectures relying on either terrestrial or satellite systems have been proposed~\cite{ps5}.
All these architectures were tailored to meet the specific PPDR requirements, including the need to support a scalable network provision for challenging environments (e.g. in terms of temperature, hygrometry, landforms, obstacles, etc.), with robust equipment to be operated while the user is in motion, able to achieve a rapid deployment of a temporary infrastructure and services to address network outages in the affected zone~\cite{itu1}. In this regard, the European Electronic Communications Committee (ECC) defines broadband PPDR temporary additional capacity as the means to provide additional network coverage at the scene of the incident with equipment such as ad hoc networks \cite{cept2}. Over the last few years, airborne networking has also gained momentum to roll out rapidly deployable PPDR communication systems. In fact, aerial platforms, which can be designed to fly and operate at different altitudes, are increasingly valued for the multiple advantages they can offer in the context of disaster relief communications~\cite{uav1}. Some of the expected benefits are favorable propagation conditions with frequent line-of-sight transmissions, low latency compared to satellite equipment, and communication payload modularity enabling mission versatility.
Moreover, regulation bodies have started considering aerial platforms as a viable set of technologies to roll out an emergency response in the first hours after a disaster~\cite{ps3}, with many kinds and sizes of aerial platforms. Basically, each platform type displays distinct features best suited to different applicative perspectives: for instance, low altitude tethered balloons have been largely investigated in the context of disaster relief~\cite{ps1},~\cite{ps2},~\cite{ps4} for their cost-efficiency, low complexity of operation and ability to rapidly lift a telecommunication payload and act as temporary cell towers as long as response operations are required. Other initiatives investigate the use of high altitude, long endurance platforms in the context of disaster relief communications~\cite{ps6}. Although still facing multiple technical challenges, such solutions should provide large network coverages from the low stratosphere during weeks or months. 

In this work, we take a particular interest in low altitude platforms with high mobility dynamics, also known as unmanned aerial vehicles, for their aptitude to self-propel and quickly bring small telecommunication payloads where and when needed. Further, we consider autonomous and distributed mobility control mechanisms for these platforms: those mechanisms, when they also enable neighboring nodes to cooperatively adapt their respective trajectories and behavior via local information exchange, are known under the name of swarming (or flocking) strategies~\cite{swarm1},~\cite{swarm2}. In this regard, we investigate how to design an efficient swarming strategy based on virtual force principles, with the objective to deploy steerable nodes that are able to form a temporary wireless network and support disaster relief communications. This paper is organized as follows: Section~\ref{sec:scenario} presents the disaster relief scenario as well as the network topology that are considered in the context of this study. Section~\ref{sec:design} details prominent works and principles related to virtual force-based mobility control mechanisms and explains the design choices made for the extended Virtual Force Protocol (VFPe), which builds upon our previous study~\cite{vfp2} by generalizing the force system and enabling node location dissemination in the network via extended multi-purpose beacon messages. Further, we analyze in Section~\ref{sec:performance} the performance of VFPe through representative network simulations, and finally conclude in Section~\ref{sec:conclusion}.

% ===========
% Section: scenario 
% ===========
\section{Scenario}
\label{sec:scenario}

\begin{figure}
\centering
\includegraphics[height=5cm]{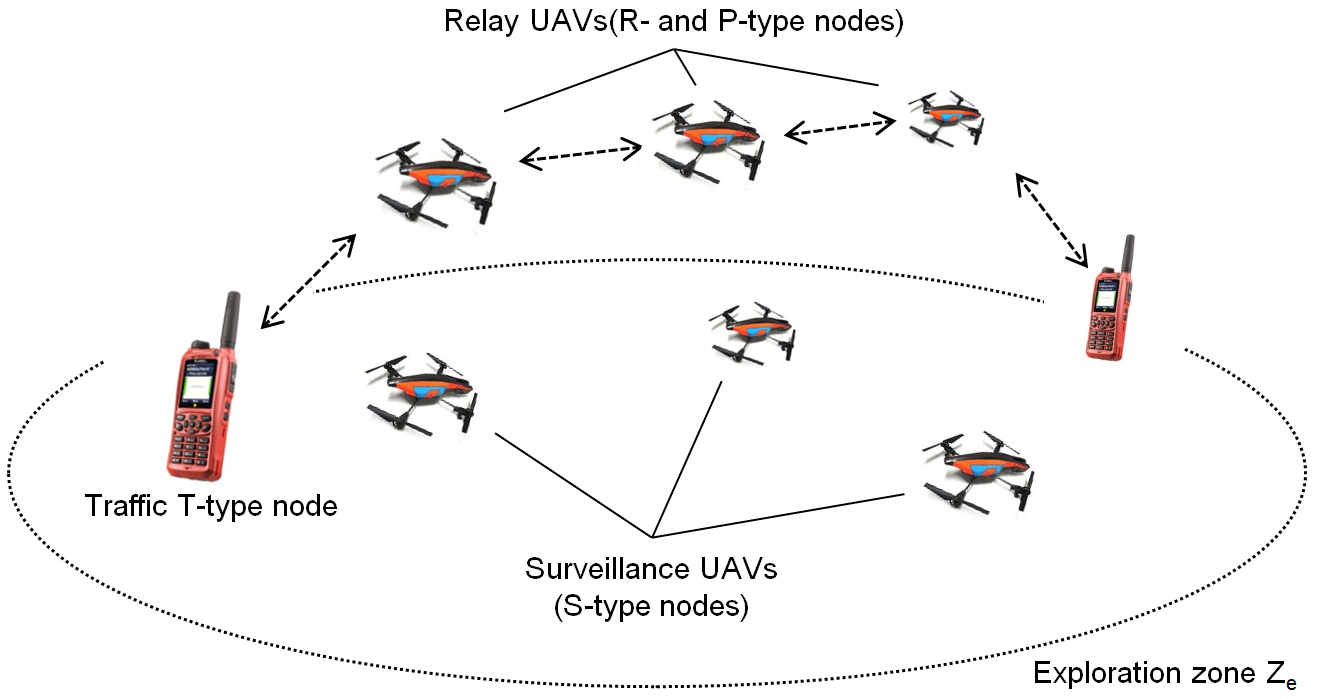}
\caption{Outline of the network equipment in the considered Disaster Relief scenario.}
\label{fig:topology}
\end{figure}

Regarding disaster relief scenarios, the main chronology events generally encompass three distinct phases: preparedness, response and recovery. We propose a network architecture which targets the response stage in this time line, with the specific objective to offer as quickly as possible a reliable and efficient communication environment to the public protection staff, rescue teams and other end users located at the scene of the considered incident or emergency. To this end, we specify four types of network nodes, as Fig.~\ref{fig:topology} shows:

\begin{itemize}
\item \emph{Traffic (T-type)} nodes impersonate consumer devices and subsequent traffic requirement on the disaster site.  A pair (source, destination) is arbitrarily chosen from this set and in the subsequent simulations, we monitor the user traffic from source to destination.
\item \emph{Surveillance (S-type)} nodes roam the exploration zone $Z_e$, discover nodes in physical proximity and record their location for later use and dissemination.
\item \emph{Relay (R-type)} nodes are former S-type nodes which became intermediate nodes in a communication chain between a (source, destination) T-type pair. When no longer useful in the chain, those nodes can revert to the S-type.
\item A \emph{prospection (P-type)} node is physically located at the apex of a forming communication chain. It not only has the role of intermediate node like regular R-type nodes, but is also in charge to evaluate which neighboring S-type node should be changed into a R-type node to further extend the communication chain. When no longer legitimate in this role, a P-type node switches to the R-type or S-type, depending on the context.
\end{itemize}

Regarding node movements, two mobility patterns can be distinguished:
\begin{itemize}
\item S- and T-type nodes roam the exploration zone $Z_e$ in a similar fashion, at random, however T-type nodes possess a slower pedestrian-like velocity compared to the other nodes.
\item R- and P-type nodes enforce a cooperative mobility strategy which we present and study in this work. With this mobility pattern, the considered R- and P-type nodes exert mutual virtual forces which result into node movement.
\end{itemize} 
These mobility schemes are further detailed in the rest of this study, and nodes are in particular given specific mobility patterns in Section~\ref{sec:performance}. Moreover, in terms of network deployment, we explain in the next section how nodes use the VFPe protocol to cooperatively create a communication chain between a user traffic (source, destination) pair through local information exchange.

% ===============
% Section: protocol design 
% ===============
\section{Protocol design}
\label{sec:design}
\subsection{Related works}
With regard to swarming concepts applied to network communications, two prominent approaches have been pursued, based on either the principles of stigmergic collaboration or physicomimetics: 

- Stigmergy, dating back to studies on social insects performed in the late 1950s~\cite{stigmergy}, refers to the ability, for a swarm of nodes, to adopt an emergent behavior via cooperation and traces left in the environment, with the example of virtual pheromones. These messages intrinsically embed the location of their emission and may allow the creation of gradients based on pheromone additivity and decay~\cite{swarm3}. Stigmergy can therefore be used to design path planning or obstacle avoidance strategies and more generally support collective node behavior. Yet, the implementation of pheromones requires either a centralized entity to act as the environment or an exchange system of incomplete views of the environment, which can both prove impractical in actual deployments. 

-  Physicomimetics, also known as physics-based swarm intelligence, applies to controlled mobility principles for which the local interactions between neighboring nodes allow nodes to reach an emergent cooperative behavior~\cite{vfp3}. In this respect, many works investigated the use of virtual forces, often derived from analogies with gravitational or electromagnetic forces~\cite{vfp3},~\cite{vfp2},~\cite{vfp5} or other physics-based phenomena~\cite{vfp4}. Despite differences in the way virtual forces are defined, such solutions share a similar approach in the way nodes evaluate, via local sensing or information exchange, the resulting virtual forces exerted by their neighbors.

In this study, our Virtual Force Protocol design proposal also relies on the principles of physicomimetics. However, the general approach in the literature related to virtual forces presupposes swarms made of a large number of nodes, and incurs the formation of large-scale and steady topologies such as vast grids~\cite{vfp3}, lattices or hexagonal distributions~\cite{vfp5}. Recent works investigate more dynamic and interest-driven topology formations~\cite{vfp4} yet still address mesh topologies with many redundant communication links, best suited for mobile sensor networks. In contrast, we seek to address deployment scenarios in which the number of network nodes is constrained. In this challenging environment, our solution can steer a limited number of relay nodes to form suitable multihop routes between nodes in need to establish a communication. To meet this requirement, two classes of VFPe core mechanisms were designed: i)  a virtual force system exerted on all intermediary nodes of a communication chain, providing the means to control their mobility pattern and ii) a beacon-based mechanism to create and maintain communication chains made of P- and R-type intermediary nodes between the user traffic endpoints. Beacons are also used to discover neighboring nodes and disseminate their positions, thereby allowing chain nodes to accurately match the desired topology. The following subsections detail both mechanisms.

\subsection{Virtual force-based system}

With VFPe, three types of virtual forces are applied on the network nodes. The first two forces are based on physicomimetics precepts~\cite{vfp3}, respectively impersonating physics-based attraction-repulsion and friction forces. As specified in our earlier works  ~\cite{vfp1},~\cite{vfp2}, both forces can be used to maintain neighboring nodes at a required distance by means of local distance-based interactions and without the need to interact with a centralized mobility planning entity.

\begin{figure}
\centering
\includegraphics[height=6.2cm]{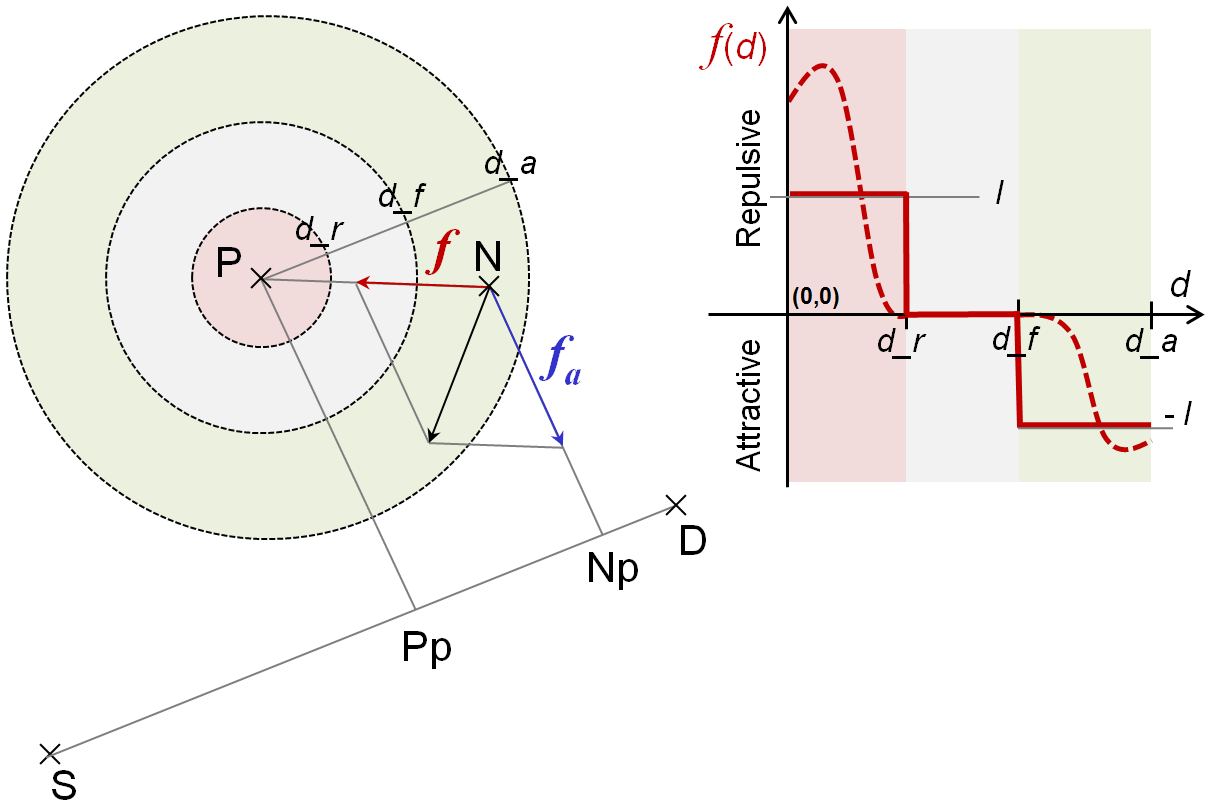}
\caption{Virtual forces $\vec{f}$ (attraction/repulsion) and $\vec{f_a}$ (alignment) exerted on a node.}
\label{fig:forces}
\end{figure}

~~\\
\textbf{An attraction-repulsion force $\vec{f}$} is defined so that any node N located in either the attractive or repulsive zone of another node P is under the influence of P's force $\vec{f}$, collinear with vector $\vec{PN}$, as illustrated by Fig.~\ref{fig:forces}: P's repulsive zone is the disc centered on P with a radius $d\_r$, while P's attractive zone is the annular surface positioned at a distance $d\_f \leq d \leq d\_a$ from P. Further, $\vec{f}$'s intensity can be defined so that it depends on the distance between N and P, Fig.~\ref{fig:forces} outlining two exemplary intensity profiles for $\vec{f}$. In the context of this work, we consider the simple piecewise-defined function such that $f=\|\vec{f}\|=I$ in the repulsive area, $f=-I$ in the attractive area and $f=0$ otherwise. 

~~\\
\textbf{A friction force $\vec{f_{fr}}$} is exerted on a node which is located in the friction zone of a nearby node. Retaining the previous notations, P's friction zone is the annular surface positioned at a distance $d\_r \leq d \leq d\_f$. If N were located in P's friction zone (which is not the case in the example of Fig.~\ref{fig:forces}), a friction force $\vec{f_{fr}}$, collinear with N's velocity vector, would be applied on N, so that  $f_{fr}=\|\vec{f_{fr}}\|=-Cx$. The use of a friction zone here is essential since it allows a neighboring node to decelerate and stop its motion within the boundaries of the considered zone. In this context, its absence would result into undesirable oscillating motion patterns for VFPe-enabled nodes, and be detrimental to the formation of the target chain topology.

It is worth highlighting that in ~\cite{vfp2}, we conducted a in-depth study of the aforementioned parameters $I$ and $Cx$ and in particular sought optimal values. Consequently, key VFPe parameters in Section~\ref{sec:performance} were valued accordingly. However, the referenced study considered specific deployment environments in which the position of the user traffic destination is known and where communications chains are formed in a straight line. In contrast, as previously delineated by Section~\ref{sec:scenario}, our current work addresses more general deployment scenarios where no assumption is made on the knowledge regarding the location and movement patterns of the user traffic sources or destinations. In this challenged environment, the simple combination of attraction-repulsion and friction forces only ensures an acceptable distance between successive intermediate nodes in a communication chain, but does not guarantee that those chains are correctly directed towards the traffic endpoints. We therefore added a third component in the VFPe virtual force system.

~~\\
\textbf{An alignment force $\vec{f_{a}}$} is used to ensure that in a communication chain, successive intermediate nodes actually direct the chain towards the user traffic destination. As again depicted by the left part of  Fig.~\ref{fig:forces}, force $\vec{f_{a}}$ tends to steer an intermediate node N of a communication chain towards the line between the user traffic endpoints S and D. Since all intermediate nodes are under the effect of $\vec{f_{a}}$, the whole chain tends to eventually form a straight line between S and D. Note that normally, $\vec{f_{a}}$ steers N towards its orthogonal projection Np on line (SD), unless this position is farther away from D, compared to the projection Pp of N's predecessor, P, on (SD). In this latter case, to make sure that N is closer from D compared to its predecessor P, $\vec{f_{a}}$ steers N towards the symmetric point of Np about Pp on (SD). That way, $\vec{f_a}$ not only allows aligning and orienting a chain towards its user traffic destination, but also triggers the repositioning of intermediate nodes if their order in the chain does not match their respective distance with destination D.

\subsection{Beacon-based protocol design}
A realistic implementation of a distributed force-based scheme requires the local exchange of information between neighboring nodes, which can be achieved via the emission by each node at regular time intervals of a specific beacon message over the radio communication links. With VFPe, we designed a beacon containing information about the emitting node, in order to enable the creation and maintenance of communication chains between user traffic sources and destinations~\cite{vfp2}. In essence, the beacon entries relate to the emitting node identifier and type, position and velocity vectors. Also, if the considered node is a relay in a chain, it contains information about successor, predecessor, destination identifiers and whether a new nearby node should be inserted in the chain as an additional relay. On this basis, VFPe is able to contextually insert S-type nodes into a chain, or on the contrary remove intermediary P- or R-type nodes, which then revert back to their S-type state. 

\begin{figure}
\centering
\includegraphics[width=10cm]{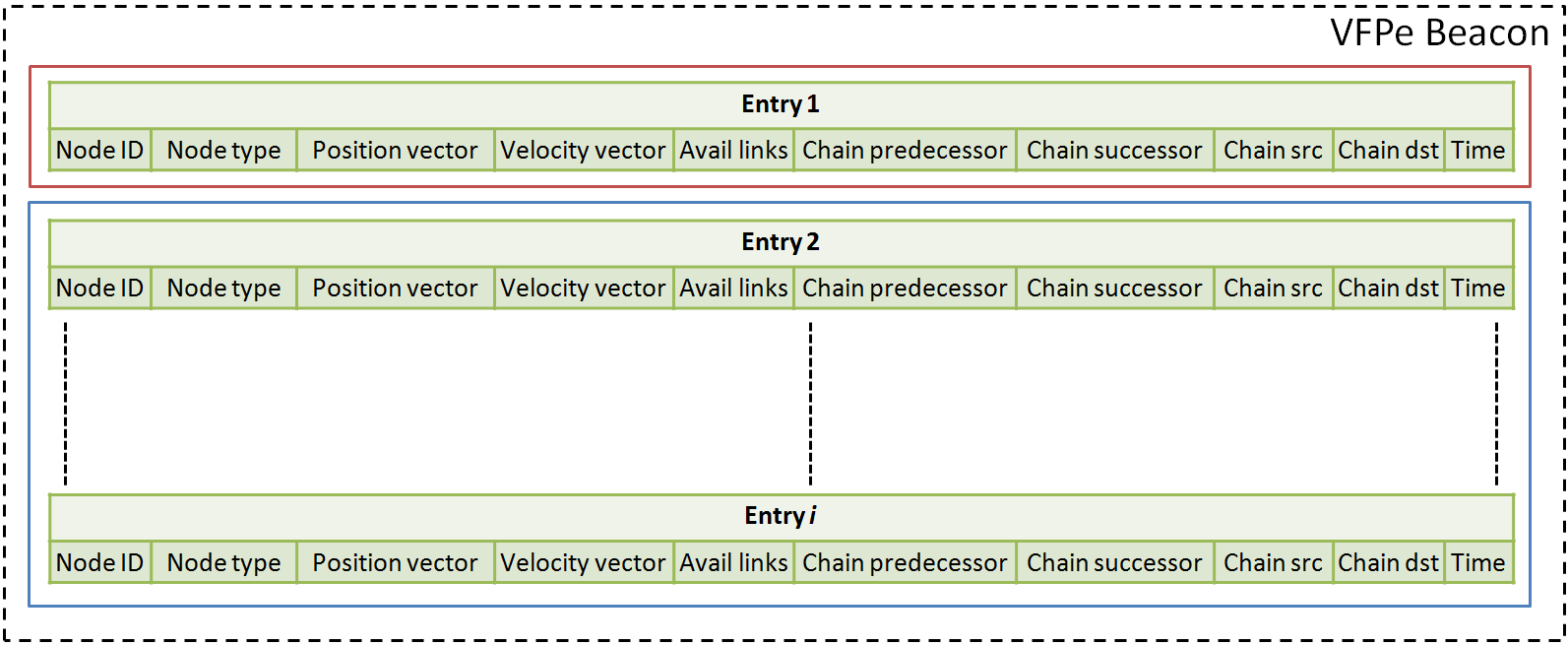}
\caption{Multi-entry structure of a VFPe beacon}
\label{fig:beacon}
\end{figure}

Compared to ~\cite{vfp2}, we extended the VFPe beacon specification in order to meet our PPDR scenario requirements where the position of user traffic endpoints is initially unknown, must be discovered, and varies with time. Our additions are twofold: 

\begin{itemize}
\item Supplementary fields were added to account for both source and destination position discovery. A time-related field was also inserted to time-stamp the whole entry and  be able to assess position errors with respect to time, as well as to allow the implementation of a VFPe scheme where beacons optionally use the freshest entries, as will be seen in the next section.
\item As illustrated by Fig.~\ref{fig:beacon}, the VFPe structure was extended to allow multiple node entries, the first entry always relating to the beacon emitting node. Additional entries are used to more quickly disseminate network node position and status. The number of entries in the beacon as well as the way entries are sorted is tunable, and is investigated further in the rest of this work.
\end{itemize} 

% ====================
% Section: performance evaluation
% ====================
\section{Performance evaluation}
\label{sec:performance}

\subsection{Simulation parameters}

As previously mentioned, VFPe entails both an exploration phase as well as a chain formation and maintenance phase. Since we were particularly interested to evaluate the impact of the exploration schemes on the overall VFPe performance, we designed the set of simulations accordingly. All (P, R, S and T)-type node mobility schemes were implemented within the release 3.23 of the ns-3 network simulator. Network nodes embed an IEEE 802.11b/g network interface card configured for High-Rate Direct Sequence Spread Spectrum (HR-DSSS) at 11 Mb/s. Moreover, the Optimized Link State Routing protocol (OLSR)~\cite{olsr} supports here multi-hop communication capability, and lossless radio propagation are assumed for the communication devices. Two T-type nodes constitute a user traffic (source, destination) pair and are initially scattered at random within the exploration area $Z_e$. Other nodes start with the S-type status and are initially located at the center of $Z_e$. Moreover, regarding the controlled mobility of the R-type and P-type nodes, VFPe is configured so that each node exerts a virtual force with key parameters valued as shown in Table I. On this matter, it is worth noting that the description of these force-based parameters, as well as the justification for the chosen values, which allow an efficient use of the VFPe protocol, can be found in~\cite{vfp2}. Likewise, mobility patterns are valued in Table I, and the Random Waypoint Model (RWP) is mentioned when relevant.

\begin{table}[h]
\begin{center}
\caption{Main simulation parameters}
\begin{tabular}{| p{2.6cm}|p{2.2cm}| p{6.8cm}|}
\hline
Exploration zone	& \multicolumn{2}{l|}{$Z_e =$ 1000 m $\times$ 1000 m}\\
\hline
Mobility patterns 	& T-type 	& Position initially uniformly distributed on $Z_e$,\\
			&	 	& RWP, velocity $\in [0.25, 1]$ m/s\\
\cline{2-3} 		& S-type	& Position initially at center of $Z_e$,\\
 			&	 	& RWP, velocity $\in [5, 10]$ m/s\\
\cline{2-3} 		& P- and R-type	& VFPe-based mobility, velocity $\in [0, 10]$ m/s\\ 
\hline
Nodes 		& \multicolumn{2}{l|}{Number of T-type nodes = 2}\\
			& \multicolumn{2}{l|}{Number of (P+R+S)-type nodes $= N$, node mass $=1$ kg}\\
			& \multicolumn{2}{l|}{When fixed, $N = 15$. Otherwise, $1 \leq N \leq 30$.}\\
\hline
Network 		& \multicolumn{2}{l|}{802.11b/g, HR-DSSS at 11 Mb/s, radio range = 100 m,}\\
			& \multicolumn{2}{l|}{constant speed propagation delay model}\\
\hline
Routing		& \multicolumn{2}{l|}{OLSR protocol with default parameter values~\cite{olsr}}\\
\hline
VFPe parameters	& \multicolumn{2}{l|}{Beacon emission interval = 1 s}\\
\cline{2-3}\textcolor{white}{VFPe} forces	& Interaction $\vec{f}$ 	& $d\_r$ = 50 m, $d\_f$ = 75 m, $d\_a$ = 100 m, $Th\_dmin$ = 40 m, $Th\_dmax$ = 75 m, $I$ = 1 N~\cite{vfp2}\\
\cline{2-3}		& Friction $\vec{f_{fr}}$ 	& $Cx$ = 2~\cite{vfp2}\\
\cline{2-3}		& Alignment $\vec{f_a}$ 	& $f_a=2$ N if node is getting closer to the target realignment position, $f_a=4$ N otherwise\\
\hline
User traffic 		& \multicolumn{2}{l|}{CBR flow at 10 Kb/s between the T-type node pair,}\\
			& \multicolumn{2}{l|}{CBR packet size = 100 bytes}\\
\hline
\end{tabular}
\end{center}
\label{table:para}
\end{table}

Furthermore, each point of the simulated performance curves illustrated and analyzed in the rest of this section results from averaging 2000 independent simulations. The same applies for the measure of contact times between nearby S-type nodes mentioned hereafter. Additionally, the handling of error margins is such that confidence intervals are constructed at a confidence level of 95\%. Error bars are shown accordingly for each performance curve in the subsequent figures. Finally, results are analyzed based on the following performance metrics: 

\begin{itemize}
\item \emph{Packet Delivery Ratio} (PDR) is defined here with respect to the user traffic flow exchanged between the (S, D) T-type node pair. In that regard, we model this traffic with a constant bitrate (CBR) flow at 10 Kb/s, which is relevant when considering added-value, narrow band PPDR traffic such as predefined status messages and short messages, point-to-point voice communications, vehicle status and transfer of location information, and access to databases in small volumes~\cite{itu1}. More precisely, the PDR is here defined by the ratio of the number of CBR packets received by destination node D over the number of CBR packets sent by source node S.
\item \emph{End-to-end delay} measures the delay difference between the time of reception by node D of the CBR packets at the application layer and the time of emission of these packets by node S, still at the application layer.
\end{itemize} 

\subsection{Simulation results}
We first took interest in how the performance of the controlled mobility-based VFPe protocol evolves with the way VFPe beacon messages are disseminated by the S-type nodes, when roaming the exploration area $Z_e$. As described in Section ~\ref{sec:design}, VFPe-enabled nodes regularly emit a beacon message, which contains a maximum of $cs$ entries related to previously discovered nodes. Besides the first entry which is related to the emitting node itself, we purposely did not specify at the design stage how the remaining entries should be chosen and prioritized by the emitting node. Instead, we implemented 3 variants of the VFPe protocol, in terms of how VFPe-enabled nodes broadcast their beacon messages:
\begin{itemize}
\item \emph{VFPe with random contacts}: according to this scheme, a VFPe-enabled node chooses the remaining entries at random, irrespectively of the time elapsed since the local information related to these nodes was first generated.
\item \emph{VFPe with fresh contacts}: this strategy, unlike the previous scheme, selects the remaining entries to insert in the VFPe beacon by prioritizing the freshest entries (i.e. which were added or updated last in the node local database).
\item \emph{VFPe with ideal node knowledge}: we also designed a VFPe extension in which the knowledge of other nodes' coordinates is not approximate and based on the disseminated VFPe beacons, but is assumed to be perfectly known at all times. We considered this ideal strategy to handle an upper range performance as a reference ; however, this theoretical extension does not address how, and at what cost in terms of signalization overhead, these exact positions could be realistically retrieved in an actual network deployment.
\end{itemize} 

In addition, we sought to complement the simulations by evaluating the performance of a RWP-only mobility scheme that does not make use of the VFPe controlled mobility. No VFPe beacons are in this case disseminated and therefore, S-type nodes are never turned into relay nodes. As a consequence, communication chains between the T-type (S,D) pair are never constructed and nodes, whether S- or T-type nodes, only rely on a RWP mobility pattern and on the underlying Mobile Ad Hoc Networks (MANET) OLSR routing protocol.

\begin{figure}
\centering
\begin{tabular}{cc}
\includegraphics[width=6.1cm]{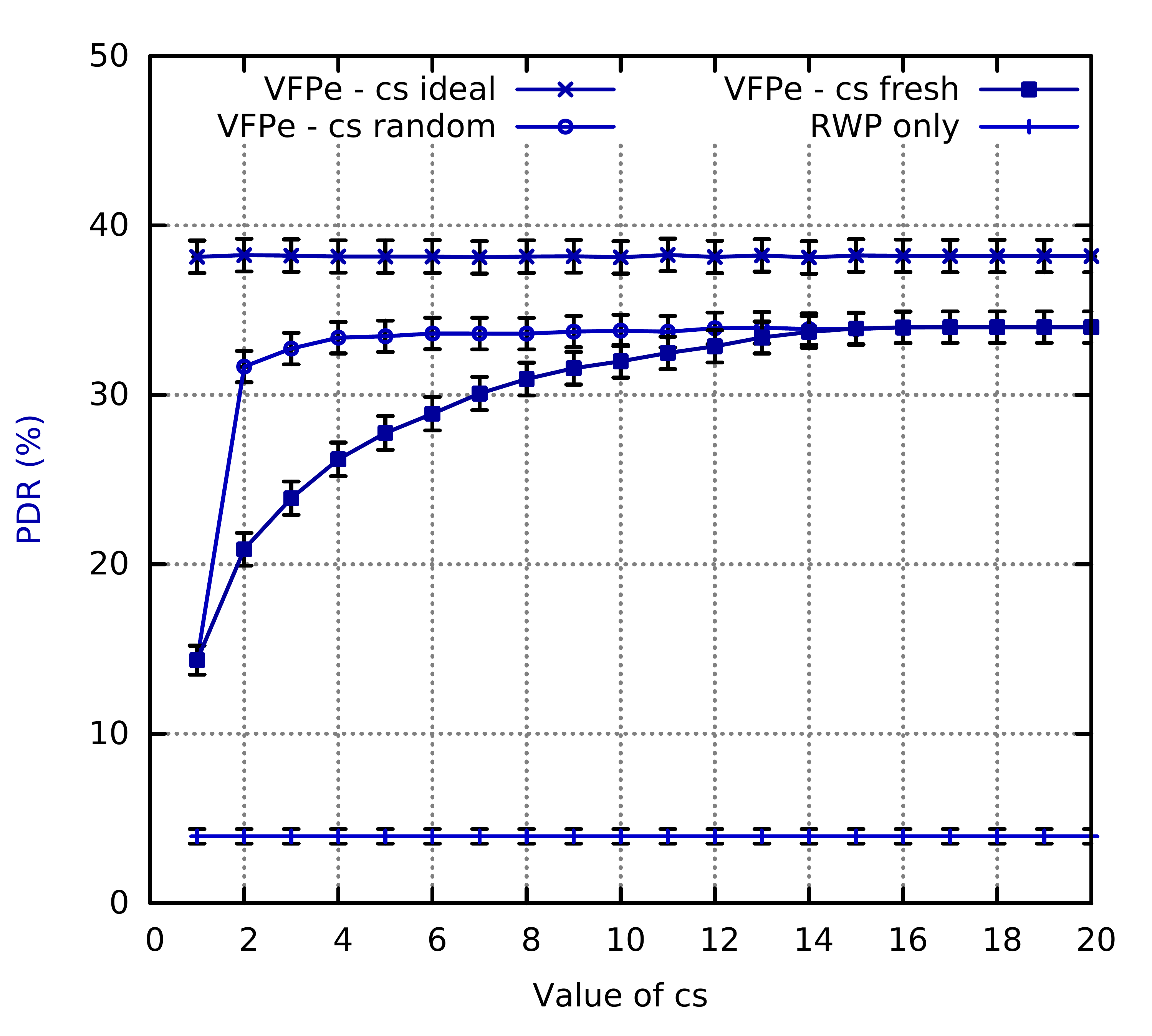} & \includegraphics[width=6.1cm]{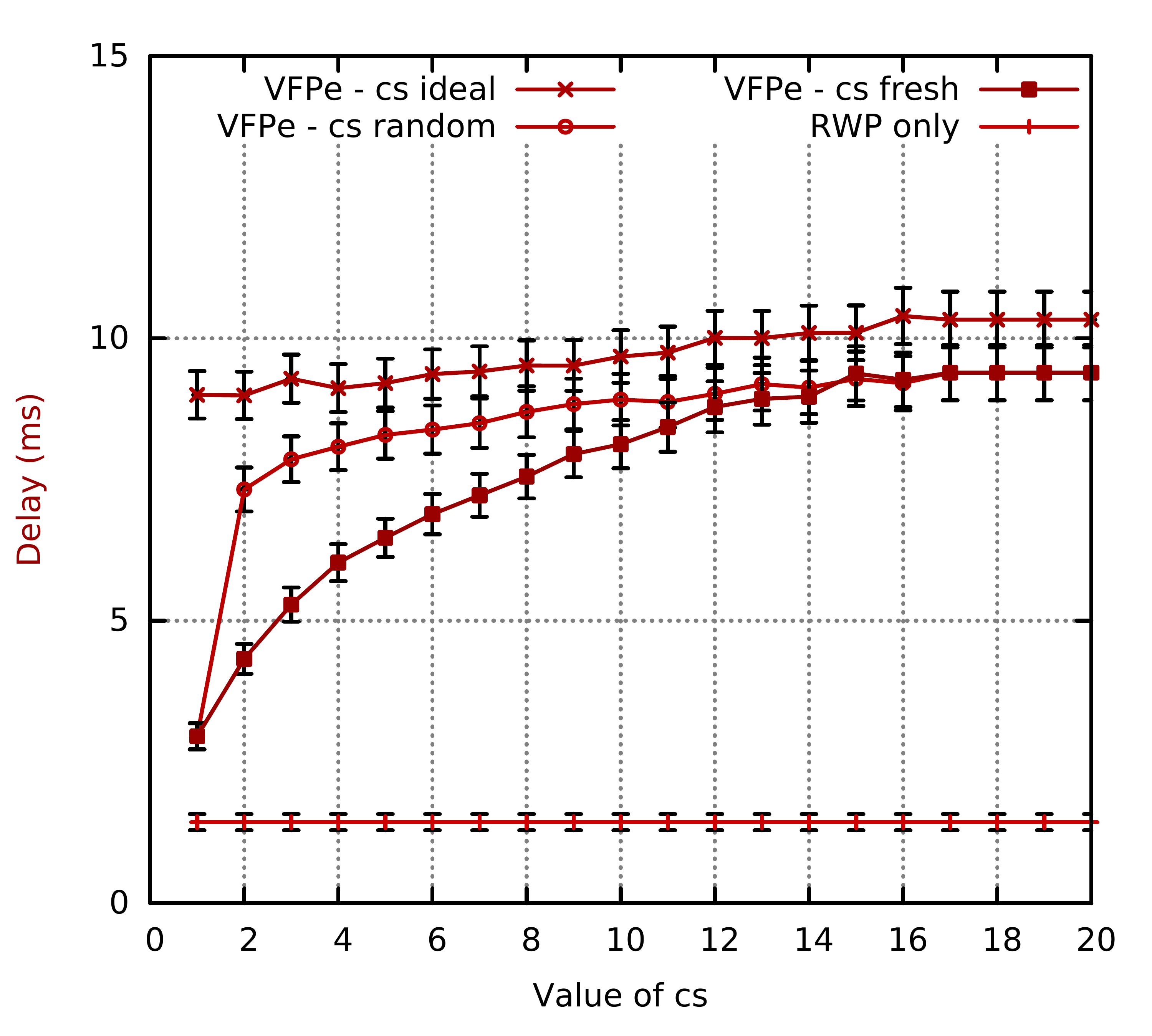}
\end{tabular}
\caption{PDR (left) and end-to-end delay (right) related to the CBR packets received by destination node D, with the number \emph{cs} of node entries contained in beacon messages.}
\label{fig:results1}
\end{figure}

~~\\
\textbf{Incidence of the contact size.}
As previously mentioned, we define the contact size as $cs$, the maximum number of entries contained by a VFPe beacon. In this first series of simulations, we studied the impact of the number of node information exchanged during a contact between nodes in direct range able to exchange VFPe beacon messages. To this end, we carried out simulations with different values $1 \leq cs \leq 20$ and measured the PDR of the CBR flow between the (S, D) T-type node pair, as well as the average end-to-end delay taken by this flow. As illustrated by Fig.~\ref{fig:results1} (left), the RWP-only and VFPe-cs-ideal schemes give a low and high performance range with a respective PDR of about 4\% and 39\%, which both remain constant with $cs$. At this stage in the analysis, it must be stressed that in the context of the considered network scenario, a maximum PDR value below 40\% does not imply here a performance issue and is the expected consequence of the incompressible time needed for unmanned vehicles to move and for the desired chain topology to be established and allow the transmission of user traffic.

Regarding the VFPe schemes with beacon messages containing either random entries (VFPe-cs-random) or fresh entries (VFPe-cs-fresh), it can be observed that the PDR is about 14\% for $cs=1$ (i.e. when nodes only disseminate their own information in the VFPe beacons) and increases with $cs$, converging towards a similar PDR value around 34\%. Further, the random scheme exhibits a significantly steeper increase, since the PDR is already approaching 32\% for $cs=2$, while the PDR of the fresh scheme is still below 22\%. Although this behavior may seem counter-intuitive at first, it can however be explained with the scenario assumptions. As an illustration, with our parameter values and setting $cs=5$, we measured for the random scheme an average contact time of 15.4 s between the nearby S-type nodes, which is much larger than the VFPe beacon emission interval. More generally, several successive beacons are likely to be exchanged during such contact times, and the random scheme will offer more diversity in terms of disseminated node entries. As a result, even with low $cs$ values, nodes will handle more accurate coordinates with the random scheme and will be steered towards positions more likely to create the desired communication chains between the (S, D) T-type node pair.

In terms of end-to-end delay, Fig.~\ref{fig:results1} (right) shows the performance of the considered four schemes. The RWP-only scheme does not use VFPe beacons, hence exhibiting a delay which remains constant with $cs$. In contrast, the VFPe-cs-ideal scheme increases with $cs$ despite a constant PDR on the whole range of $cs$, as previously seen. To explain this, it is worth highlighting that in the implementation of our 3 VFPe-enabled schemes, each node entry takes 36 bytes in a VFPe beacon message. Besides, although all simulations rely on a HR-DSSS modulation at 11 Mb/s, the VFPe frame payload is transmitted at a lower bitrate of 1 Mb/s, since the considered beacons are broadcast. As a result, each additional node entry (and therefore incrementation of $cs$) incurs an increase of about 0.29 ms, which in turns impacts the CBR queued traffic awaiting transmission on the wireless shared medium. Moreover, the delays for both VFPe-cs-random and VFPe-cs-random schemes are lower than 9.5 ms for the whole range of $cs$ values and remain compatible with the low-latency requirements of interpersonal communications, for instance. In that light, the use of the VFPe-cs-random scheme may be preferred since the PDR approaches its maximum value with only a low contact size value (e.g. $cs=4$), that is with a small-sized beacon. 

\begin{figure}
\centering
\begin{tabular}{cc}
\includegraphics[width=6.1cm]{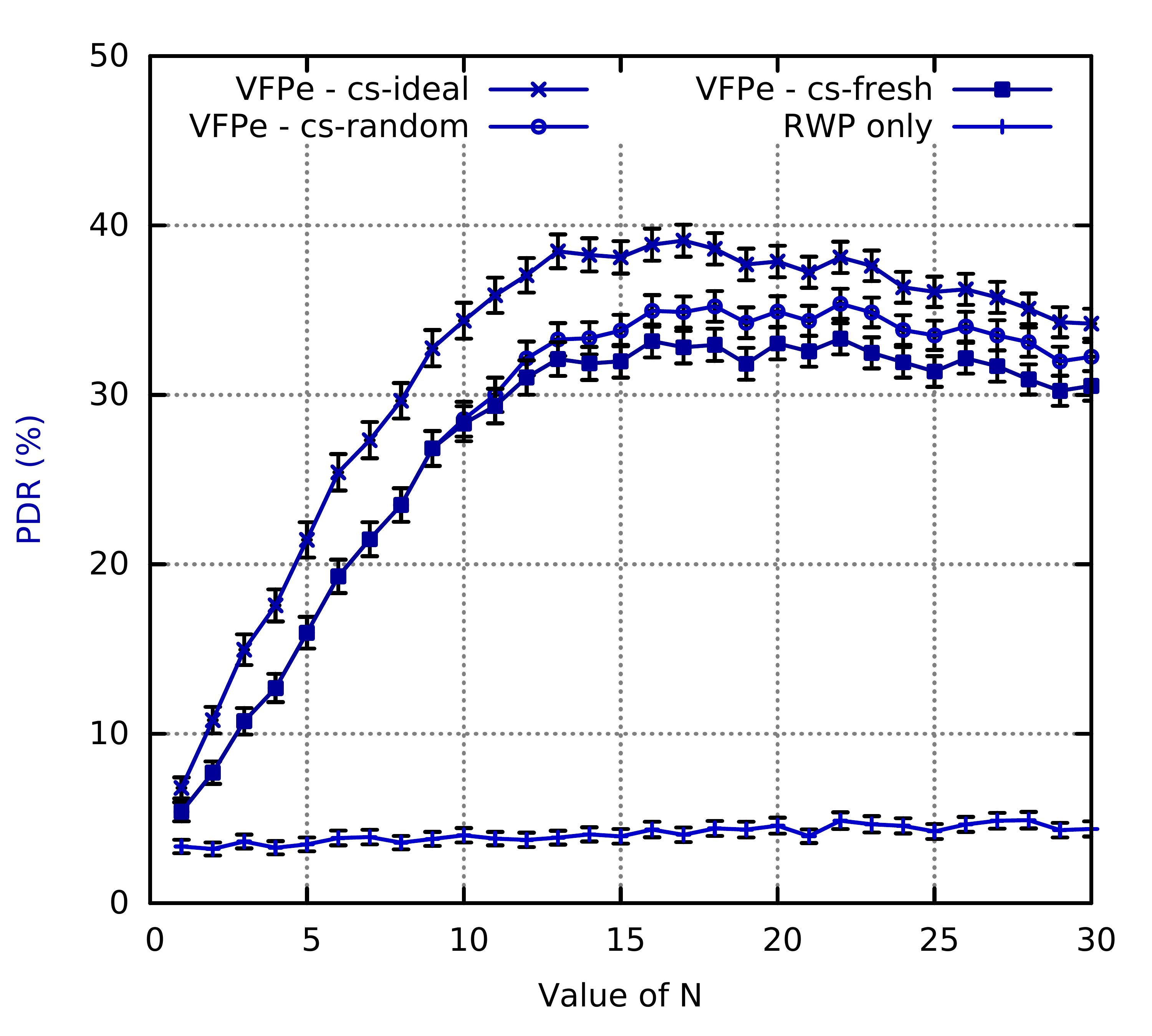} & \includegraphics[width=6.1cm]{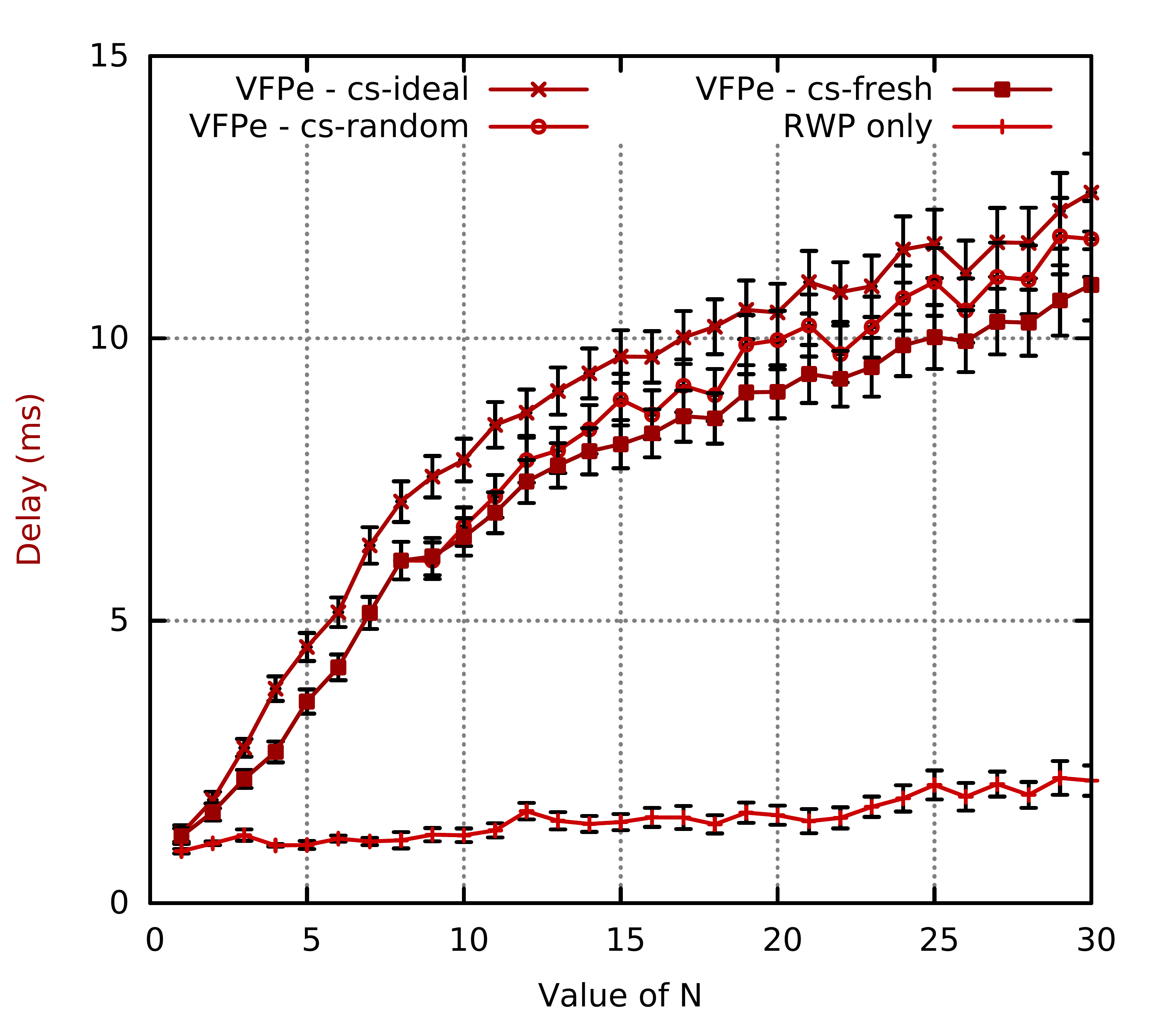}
\end{tabular}
\caption{PDR (left) and end-to-end delay (right) related to the CBR packets received by destination node D with \emph{N}, the initial number of S-type nodes.}
\label{fig:results2}
\end{figure}

~~\\
\textbf{Incidence of the number of nodes.} For any given deployment scenario, the number of rolled-out systems has a strong impact on the workability of operations and on the overall cost of the solution. We therefore thought to assess how the performance of a controlled mobility scheme like VFPe evolves with the number of network nodes, and in particular with $N$, the initial number of S-type nodes. Fig.~\ref{fig:results2} (left) gives a representation of the PDR results for each of the four considered mobility strategies, with $1 \leq N \leq 30$. In this set of simulations as well, the RWP-only scheme gives a reference, and a low performance range, regarding what can be expected from a simple MANET protocol here: overall, the PDR slightly increases with $N$, ranging from 3\% up to 4.5\%. Assuming here a constant value $cs=10$, the three VFPe-enabled schemes give PDR results which are consistent with the previous set of simulations: the ideal scheme yields the best results for all values of $N$ in the considered interval, while VFPe-cs-random outperforms VFPe-cs-fresh for $N \geq 10$. Besides, all three schemes show a steep PDR increase, then a maximum PDR value for $N$ around 16 to 18, followed by a regular PDR decrease in the rest of the considered interval for $N$.

Moreover, Fig.~\ref{fig:results2} (right) allows stressing that the end-to-end delays for the three VFPe-enabled schemes, although still acceptable since they remain below 15 ms for all considered values of $N$, sharply increase with $N$. As a whole, this set of simulations shows that a limited set $N_{max}$ of S-type nodes (such that $16 \leq N_{max} \leq 18$ with our assumptions) is needed to reach the best network performance. At this point, in order to properly apprehend those results, it is important to recall that the speed range of exploration S-type nodes, which is $[5,10]$ m/s, exceeds the standard speed conditions normally found in pedestrian-type scenarios, where MANET protocols, and in particular link-state protocol like OLSR, behave best. As a result, when $N=16$, enough S-nodes are initially available so that VFPe induces the best possible chain topologies. As a result, above that threshold for $N$, no additional S-type node is statistically expected to be converted into a relay (P, R)-type node to extend the communication chains, and the extra S-type nodes will be used by VFPe for exploration purposes only. When $N$ increases above that threshold, the beneficial effects of adding S-type nodes in the network, and therefore increasing link diversity for the underlying multi-hop routing protocol, are severely hindered by the detrimental effects of high-velocity S-type nodes crossing the communication chains. In effect, the underlying MANET protocol will detect those transient and unstable links and try to exploit the corresponding routes, which is in turn likely to result into CBR packet losses, the time for OLSR to detect the link failures and to rebuild its routes accordingly. Further, an in-depth analysis of the simulation traces confirms the increasing amount of OLSR route losses due to the temporary presence of unneeded nearby S-type nodes, when $N$ increases above the aforementioned threshold. In this regard, corrective steps could be taken to prevent the routing protocol from exploiting those suboptimal routes or even to repel S-type nodes (through an extra repulsion force effect, for instance) from an established communication chain. However, it is important to highlight that marginally improving the overall network performance via an increase of additional nodes is not desirable in the context of disaster relief deployment scenarios, for which we seek a limited number of deployed communication systems. With that requirement in mind, the implications from this analysis are threefold: first, VFPe allows reaching a satisfying network performance with a limited number $N_{max}$ of nodes, then that this number $N_{max}$ can be precisely defined, and lastly, that $N_{max}$ remains sufficiently low ($16 \leq N_{max} \leq 18$ with the given assumptions) to offer the perspective of cost-efficient network deployments in real conditions. 

% ============
% Section: conclusion 
% ============
\section{Conclusion}
\label{sec:conclusion}
In this work, we presented the VFPe protocol, which, by the use of virtual force principles, allows a swarm of mobile network nodes to fly cooperatively, to rapidly form wireless multi-hop communication routes and to efficiently transmit end-user traffic flows. After describing our network architecture proposal and explaining the design choices made for VFPe, notably in terms of force system and beacon-based information dissemination in the network, we analyzed a series of simulation results to assess the impact of the exploration schemes on the overall VFPe performance. We first analyzed how VFPe behaves with the way its beacon messages are disseminated by the network nodes in their exploration phase, and verified that our proposals for realistic VFPe scheme implementations compare satisfyingly with an ideal strategy. We then scrutinized the progression of VFPe performance with respect to the number of network nodes and identified, in the specific context of high mobility underpinning our applicative scenarios, that a maximum efficiency can be obtained with a limited set of nodes. Those results lead us to conclude that VFPe is able to provide an adapted solution for the rapid deployment of temporary networks, notably suited to the demanding context of disaster relief and cost-efficient emergency communications where in particular the amount of communication systems, and in this case the number of mobile network nodes, is constrained. Currently, we are planning to embed a VFPe implementation on quad-rotors to further evaluate the solution through experimentation. In the future, we intend to study the benefits of jointly using a virtual force-based protocol with disruption- and delay-tolerant schemes.

\end{document}